\documentclass[runningheads]{llncs}
\usepackage{graphicx}
\usepackage{amsmath}
\usepackage{caption}
\usepackage{multirow}
\usepackage{booktabs}
\usepackage{xcolor}

\usepackage{amsmath}
\usepackage{amssymb}
\usepackage{bm}
\usepackage{mathtools}
\usepackage{dsfont}
\usepackage{multirow}
\usepackage{leftidx}
\usepackage{soul}
\begin{document}
\title{$k$-$t$ NEXT: Dynamic MR Image Reconstruction Exploiting Spatio-temporal Correlations}
\titlerunning{$k$-$t$ NEXT: $k$-$t$ NEtwork with X-$f$ Transform}
\author{Chen Qin\inst{1}, Jo Schlemper\inst{1}, Jinming Duan\inst{1,3}, Gavin Seegoolam\inst{1}, Anthony Price\inst{2}, Joseph Hajnal\inst{2}, Daniel Rueckert\inst{1}}
\institute{Department of Computing, Imperial College London, London, UK \email{c.qin15@imperial.ac.uk}\\
\and
Division of Imaging Sciences and Biomedical Engineering Department, King’s College London, St. Thomas’ Hospital, London, UK\\
\and
School of Computer Science, University of Birmingham, Birmingham, UK}
%
\authorrunning{C. Qin et al.}
%
%
\maketitle              
\begin{abstract}
Dynamic magnetic resonance imaging (MRI) exhibits high correlations in $k$-space and time. In order to accelerate the dynamic MR imaging and to exploit $k$-$t$ correlations from highly undersampled data, here we propose a novel deep learning based approach for dynamic MR image reconstruction, termed $k$-$t$ NEXT ($k$-$t$ NEtwork with X-$f$ Transform). In particular, inspired by traditional methods such as $k$-$t$ BLAST and $k$-$t$ FOCUSS, we propose to reconstruct the true signals from aliased signals in $x$-$f$ domain to exploit the spatio-temporal redundancies. Building on that, the proposed method then learns to recover the signals by alternating the reconstruction process between the $x$-$f$ space and image space in an iterative fashion. This enables the network to effectively capture useful information and jointly exploit spatio-temporal correlations from both complementary domains. Experiments  conducted on highly undersampled short-axis cardiac cine MRI scans demonstrate that our proposed method outperforms the current state-of-the-art dynamic MR reconstruction approaches both quantitatively and qualitatively.

\end{abstract}
\section{Introduction}
Dynamic Magnetic Resonance Imaging (MRI) is a non-invasive imaging technique to monitor dynamic processes such as cardiac motion by acquiring data in a $k$-$t$ space that contains both temporal and spatial information. However, the acquisition speed is limited due to both physical and physiological constraints. It is well known that in dynamic MRI there exists significant correlations in $k$-space and time. In order to increase the acquisition rate, most strategies have been designed to acquire part of the desired $k$-$t$ measurements and then reconstruct the images by exploiting spatio-temporal redundancies within the data. 

Inspired by traditional $k$-$t$ methods from the area of compressed sensing \cite{jung2009k,jung2007improved,tsao2003k} for accelerated dynamic MR imaging, here we propose a novel dynamic MR image reconstruction NEtwork with X-$f$ Transform, termed $k$-$t$ NEXT, which exploits the signal redundancies in both $x$-$f$ domain and image domain. In particular, the proposed $k$-$t$ NEXT formulates the reconstruction process in an iterative fashion, where in each iteration, it consists of two sub-modules: a $xf$-CNN that learns to recover the true signals from aliased signals in $x$-$f$ domain, and a convolutional recurrent neural network (CRNN) that exploits spatio-temporal redundancies in image domain. The dynamic reconstruction process thus alternates between $x$-$f$ space and image space, which potentially enables the network to learn complementary features simultaneously from both domains. Experiments were performed on highly undersampled short-axis cardiac cine MR scans, where we show that the proposed model outperforms the current state-of-the-art dynamic MR reconstruction methods. 

\subsection{Related Work}
Over the years, a number of approaches have been proposed for the reconstruction of accelerated dynamic MR images. In general, these methods can be mainly divided into three categories, based on exploiting correlations in $k$-space, in time, and in both $k$-space and time \cite{tsao2003k}. 
The first class of approaches exploit the correlations between $k$-space points at the same time frame, and then reconstruct each frame independently from other time frames, such as reduced field-of-view (FOV) \cite{hu1994reduction} and parallel imaging methods \cite{griswold2002generalized}, while the second group of strategies is to exploit redundancies in time, where the missing data at a given position can be interpolated or extrapolated from the measured data at other time points, such as keyhole imaging \cite{jones1993k} and data sharing \cite{zhang2010magnetic}. Relevant to our method, the third type of approaches is based on exploiting correlations in both $k$-space and time. One of the examples is the model-based $k$-$t$ BLAST and $k$-$t$ SENSE method \cite{tsao2003k}, which takes advantage of a-priori information about the $x$-$f$ support obtained from the training stage and then to remedy the aliasing artefacts during acquisition stage. Based on that, $k$-$t$ FOCUSS \cite{jung2007improved,jung2009k} then formulated the problem in a compressed sensing MRI framework, which enforced the sparsity in $x$-$f$ domain for the signal recovery. Similarly, a low rank and sparse reconstruction scheme ($k$-$t$ SLR) \cite{lingala2011accelerated} was proposed to exploit correlations between the temporal profiles of the voxels by introducing non-convex spectral norms and spatio-temporal total variation norm. 
In more recent years, deep learning approaches have gained their popularity for MR image reconstruction \cite{eo2018kiki,qin2019convolutional,schlemper2018deep,ye2018deep}. Most approaches investigate on exploiting information in a single frame (or static image) either in image domain \cite{hammernik2018learning,liu2018image,schlemper2018bayesian} or in $k$-space domain \cite{akccakaya2019scan,han2018k,zhang2018multi}, where each frame (or image) is reconstructed independently. In order to exploit the temporal redundancies, Schlemper et al. \cite{schlemper2018deep} proposed a data sharing (DS) layer in an image space cascaded 3D convolutional network to utilise the similar information contained in neighboring $k$-space samples. Qin et al. \cite{qin2019convolutional} also proposed a bidirectional CRNN model to exploit the temporal dependencies of dynamic sequences in image domain. In contrast, our approach proposes to reconstruct the images in both $x$-$f$ and image domains, where complementary information from two different domains can be fully exploited.

\section{Methods}
\subsection{Problem Formulation}
Consider a Cartesian $k$-space trajectory where $k_x$ denotes the phase encoding direction, $k_y$ denotes the readout direction, while $\sigma(x,t)$ denotes the image domain content at $x$ and time $t$. The $k$-space measurement $v(k,t)$ is then formulated as:
\begin{equation}\label{eq1}
    v(k,t)=\int \sigma(x,t) e^{-j2\pi kx}dx = \int \int \rho(x,f)e^{-j2\pi (kx+ft)}dx\,df,
\end{equation}
where $\rho(x,f)$ is the 2D spectral signal in $x$-$f$ domain. This can also be represented in a matrix form: ${\bf{v}} = \mathcal{F} {\bf{\rho}}$,
in which $\bf{v}$ and $\bf{\rho}$ stand for the stacked $k$-$t$ space measurement vectors and $x$-$f$ image respectively, and $\mathcal{F}$ is the 2D Fourier transform along the $x$-$f$ direction. From the perspective of compressed sensing, the problem can be formulated by exploiting the sparsity of the unknown signal:
\begin{equation}
\text{min} \ ||{\mathbf{\rho}}||_{1}, \quad s.t.\ ||\mathbf{v}-\mathcal{F} \mathbf{\rho}||_{2}\leq \epsilon,
\end{equation}
where $\epsilon$ denotes the noise level. In k-t FOCUSS \cite{jung2007improved,jung2009k}, the underdetermined inverse problem was solved via a sparse reconstruction algorithm called FOCUSS. The solution then can be expressed as the form that consists of a baseline signal $\bar{\rho}$ and its residual encoding for the $n$-th estimate of the $x$-$f$ signal $\rho^{(n)}$:
\begin{equation}\label{FOCUSS}
    \rho^{(n)} = \bar{\rho} + \text{FOCUSS}(\rho^{(n-1)}-\bar{\rho}, \rho^{(n-1)}).
\end{equation}
Here the mathematical form of FOCUSS algorithm is omitted for simplicity. For details, please refer to \cite{jung2007improved,jung2009k}.

\subsection{$k$-$t$ NEXT for Dynamic MRI Reconstruction}
\begin{figure*}[!t]
\centering
\includegraphics[width=0.95\linewidth]{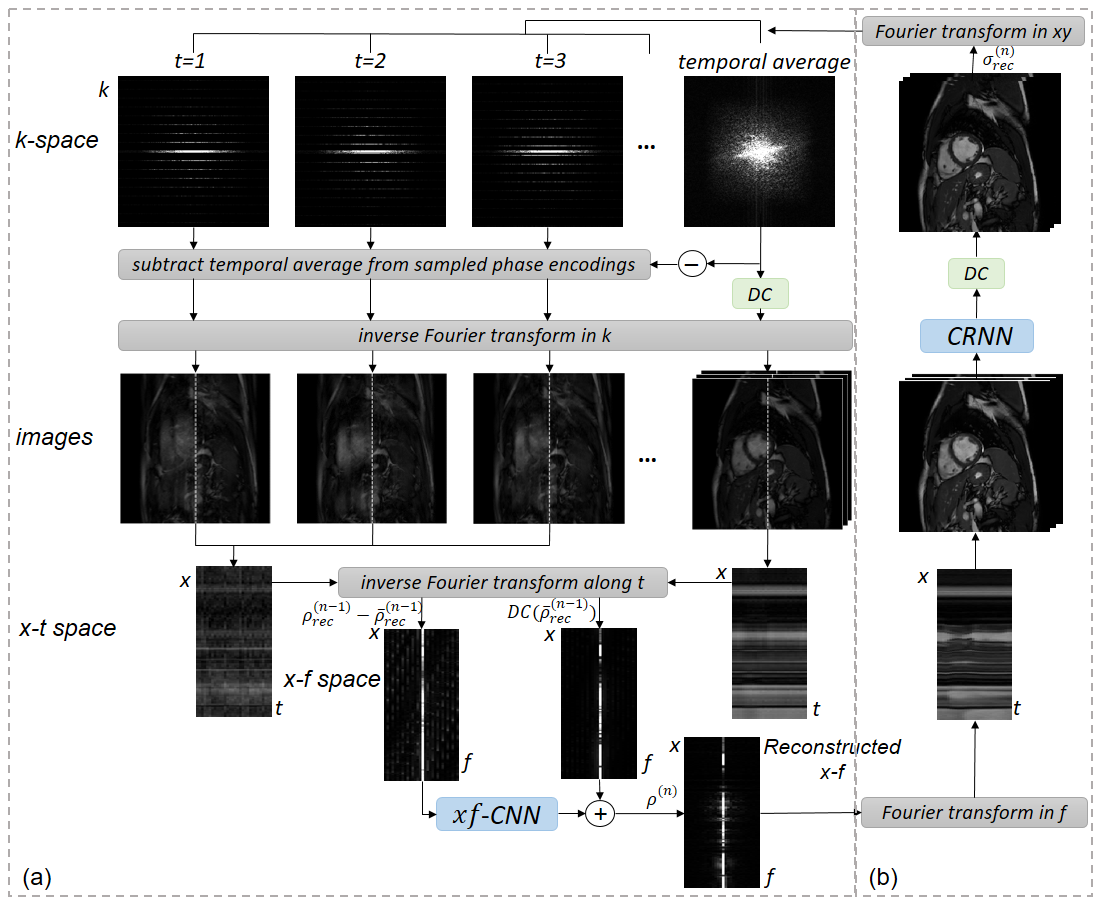}
\caption{The $k$-$t$ NEXT reconstruction diagram. True signals can be recovered by iteratively updating the reconstruction in both (a) $x$-$f$ and (b) image domains via learning the $xf$-CNN and CRNN jointly. For mathmetical notations, please refer to Eq. \ref{eq:alternate_recon}. }
\label{fig:flow}
\end{figure*}

Motivated by $k$-$t$ BLAST \cite{tsao2003k} and $k$-$t$ FOCUSS \cite{jung2007improved}, we propose a dynamic image reconstruction NEtwork with X-$f$ Transform ($k$-$t$ NEXT) to exploit the spatio-temporal correlations from both $x$-$f$ space and image space. Specifically, $k$-$t$ NEXT formulates the iterative reconstruction process in an unfolded cascading way, as it has been shown to be a powerful technique in MR reconstruction \cite{qin2019convolutional,schlemper2018deep}. In each iteration, our proposed approach learns to reconstruct the true images by alternating between $x$-$f$ and image spaces, so that the spatio-temporal redundancies can be jointly exploited from these two complementary domains. 
In particular, a $xf$-CNN is proposed for the recovery of signals in $x$-$f$ domain inspired by the traditional $k$-$t$ method, and a variation of the CRNN-MRI \cite{qin2019convolutional} network is adopted for the subsequent image space reconstruction. We can compactly represent a single iteration of the $k$-$t$ NEXT as follows:
\begin{subequations}
\label{eq:alternate_recon} 
\begin{align}
    \rho^{(n)} &= \text{DC}(\bar{\rho}_{rec}^{(n-1)}) + xf\text{-CNN}(\rho_{rec}^{(n-1)}-\bar{\rho}_{rec}^{(n-1)}), \label{xfcnn}
\\ 
    \mathbf{\sigma}_{rec}^{(n)}&=\text{CRNN}(\mathcal{F}_f\rho^{(n)}; \mathbf{v}^{(0)}), \quad \rho_{rec}^{(n)} = \mathcal{F}_f^H \mathbf{\sigma}_{rec}^{(n)}, \label{crnn}
\end{align}
\end{subequations}
where $\mathbf{\sigma}_{rec}^{(n)} \in \mathds{C}^D$ denotes the complex-valued reconstructed image sequence at iteration $n$, and $\sigma_{rec}^{(0)}=\sigma_u$ is the acquired zero-filled undersampled images. Here $D=D_xD_yT$, in which $D_x$ and $D_y$ are width and height of the frame and $T$ is the number of frames. $\mathcal{F}_f$ denotes the Fourier transform along $f$ dimension, and $\rho_{rec}^{(n)}$ is the $x$-$f$ spectral signal transformed from $\sigma_{rec}^{(n)}$, while $\rho^{(n)}$ stands for the intermediate reconstructed signal from $xf$-CNN. Also $\bar{\rho}_{rec}^{(n-1)}$ denotes the temporally averaged $x$-$f$ signal (see Eq. (\ref{temporal_average})), DC stands for the data consistency layer \cite{schlemper2018deep}, and $\mathbf{v}^{(0)}\in \mathds{C}^M$ ($M \ll D$)  is the acquired raw data.
An illustrative diagram of $k$-$t$ NEXT is shown in Fig. \ref{fig:flow}. We will introduce it in the following.

\subsubsection{${xf}$-CNN exploiting spatio-temporal correlations in $x$-$f$ domain.}
Following the formulation in Eq. (\ref{FOCUSS}), here we propose to formulate the $xf$-CNN reconstruction as Eq. (\ref{xfcnn}),
where instead of using model-based \cite{tsao2003k} or compressed sensing \cite{jung2007improved} algorithms to recover the true signals, we employ a stack of CNN layers to estimate the missing data based on other available points, typically within its vicinity in $x$-$f$ space. In particular, here the $x$-$f$ baseline signal $\bar{\rho}_{rec}^{(n)}$ is a temporal average of a sequence, i.e.,
\begin{equation}\label{temporal_average}
 \bar{\rho}_{rec}^{(n)}=\mathcal{F}^H \left[\sum_t \mathbf{v}^{(n)}./\text{max}(\mathbf{1}, \sum_{t} \delta(\mathbf{v}^{(n)}))\right], \quad \delta(a)=
\begin{cases}
0& a=0\\
1& a \neq 0
\end{cases}
\end{equation}
in which $\mathbf{v}^{(n)}$ is the $k$-space data that is Fourier transformed from $\sigma_{rec}^{(n)}$, and the $./$ and max operation is performed element-wise.
Thereby, $xf$-CNN learns to reconstruct residuals of each frame, which further exploits the signal sparsity.

The illustrative diagram of $x$-$f$ reconstruction is shown in Fig. \ref{fig:flow}(a). Specifically, we formulate the $k$-$t$ to $x$-$f$ transformation process as a {\bf{$\bf{x}$-$\bf{f}$ transform layer}} in the network. In details, the $x$-$f$ transform layer receives input from $k$-$t$ space data. For iteration $n$, the acquired $k$-space data is firstly averaged along $t$ to yield a temporal average (Eq. (\ref{temporal_average})), which is then subtracted from data at each time frame. To ensure data fidelity for the baseline estimate, here we propose to incorporate a data consistency (DC) term for $\bar{\rho}_{rec}^{(n-1)}$ at each frame separately. Then the subtracted data and temporally averaged data are inverse Fourier transformed to image space to obtain a sequence of aliased images and a data-consistent temporally averaged sequence. Each frequency-encoding position is then processed separately hereafter. The image columns from aliased images or baseline images are then gathered and inverse Fourier transformed along $t$ to yield an $x$-$f$ image, corresponding to $\rho_{rec}^{(n-1)}-\bar{\rho}_{rec}^{(n-1)}$ and DC($\bar{\rho}_{rec}^{(n-1)}$) respectively, which are then fed as inputs to $xf$-CNN for $x$-$f$ space reconstruction (Eq. (\ref{xfcnn})). After the signal de-aliasing in $x$-$f$ domain, another Fourier transform along $f$ is adopted to transform the estimated $x$-$f$ signal $\rho^{(n)}$ back to dynamic image space for the subsequent image space reconstruction (Eq. (\ref{crnn})).

\subsubsection{$k$-$t$ NEXT exploiting spatio-temporal redundancies in complementary domains.}
Previous approaches \cite{eo2018kiki} have shown that exploring cross-domain knowledge is beneficial for MR reconstruction task. Inspired by this, with the aim of exploiting redundancies in complementary domains, here we propose to learn a dynamic MR reconstruction network in both $x$-$f$ and image spaces jointly. In particular, we employ the CRNN model for image space reconstruction due to its effectiveness in exploiting temporal redundancies with a relatively smaller network capacity \cite{qin2019convolutional}. Thus, in each cascade, the proposed $k$-$t$ NEXT consists of a $xf$-CNN and a CRNN block, where it employs all 2D convolutions across spatial and temporal dimensions, in contrast to 3D convolutions used in the baseline method \cite{schlemper2018deep}. This enables the network to be more efficient and effective in learning useful and complementary features in $x$-$f$, spatial and temporal space simultaneously.

Given the training data $S$ with undersampled data as input and fully sampled data as target, i.e., $({{\bf{\sigma}}_u},{{\bf{\sigma}}_t})$ in image space and $({\bf{\rho}}_{u}, \rho_{t})$ in $x$-$f$ space, the network is trained end-to-end by minimising the pixel-wise mean squared error (MSE) between the reconstructed data and the ground truth fully sampled data:
\begin{equation}
\mathcal{L}\left( \boldsymbol{\theta}  \right){\rm{ = }}\frac{1}{n_S}\sum\limits {\left(\left\| {{{\bf{\sigma}}_t} - {{\bf{\sigma}}_{rec}^{(N)}}} \right\|_2^2+\left\| {{{\bf{\rho}}_t} - {{\bf{\rho}}^{(N)}}} \right\|_2^2\right)},
\end{equation}
where ${{\bf{\sigma}}_{rec}^{(N)}}$ and ${{\bf{\rho}}^{(N)}}$ denote the predicted image and $x$-$f$ array at iteration $N$, i.e., the final output in image domain and $x$-$f$ domain respectively, $\boldsymbol{\theta}$ is the set of network parameters, and ${n_S}$ is the number of training samples.

\section{Experiments and Results}
\subsection{Dataset and Implementation Details}
The dataset used in our experiments consists of 10 fully sampled complex-valued short-axis cardiac cine MRI. Each scan contains a single slice SSFP acquisition with 30 temporal frames. The raw data has 32-channel data with sampling matrix $192 \times 190$, which was zero-filled to $256 \times 256$, and the raw multi-coil data was then reconstructed to produce a single complex-valued image. In experiments, images were transformed back to $k$-space to simulate a fully sampled single-coil acquisition. A shear grid $k$-$t$ Cartesian sampling pattern with four central lines (see Fig. \ref{fig:xf}(b)) was employed to undersample the $k$-space data to generate the undersampled input image sequences. The undersampling rate mentioned is stated with respect to the matrix size of the data, which is $192 \times 190$.

In the proposed $k$-$t$ NEXT, $xf$-CNN is composed of 5 layers of 2D CNN with a residual connection from the baseline estimate. For the CRNN model, a variation of architecture \cite{qin2019convolutional} is employed which consists of 4 layers of bidirectional CRNN, 1 layer of 2D CNN, a residual connection and a DC layer. We used dilated convolutions with kernel size $3 \times 3$ and dilation factor $(3,3)$, and the number of cascade $N$ was set to 4 for all comparison methods. {For detailed network architecture, please refer to supplementary materials.} The network was implemented in PyTorch. During training, ADAM optimiser was employed with a learning rate of $10^{-4}$. Data augmentation was performed on-the-fly, with random rotation, scaling, and elastic transformation. All evaluations were done via a 3-fold cross validation.

\subsection{Results}
In experiments, we compared our proposed approach ($k$-$t$ NEXT) with different dynamic MR reconstruction methods, including compressed sensing method $k$-$t$ FOCUSS \cite{jung2007improved}, deep learning method CRNN-MRI \cite{qin2019convolutional}, and DS+3DCNN \cite{schlemper2018deep} that incorporates data sharing (DS). To investigate the effectiveness of $xf$-CNN, an additional baseline approach is proposed which replaces all $x$-$f$ reconstruction in $k$-$t$ NEXT with DS component, termed DS+CRNN. In DS methods, we set the number of neighbouring frame as $n_{adj} \in \{0,1,...5\}$ as in \cite{schlemper2018deep}. Note that for a fair comparison with our $k$-$t$ NEXT, we modified the baseline approaches DS+3DCNN and DS+CRNN to learn the residual of a temporally averaged frame as well. Quantitative comparison results of different methods on dynamic cardiac data with undersampling rates 9 and 12 are presented in Table \ref{results_2}, where it compares the network capacity per cascade, peak-to-noise-ratio (PSNR), structural similarity index (SSIM) and high frequency error norm (HFEN) \cite{qin2019convolutional}. Networks for different undersampling factors were trained separately in this case. It can be seen that our proposed $k$-$t$ NEXT can outperform other baseline methods by a large margin in terms of all these measures at different undersampling rates, with roughly the same level of network capacity. In particular, $k$-$t$ NEXT performs better than its corresponding DS pair, which indicates the merits of exploiting correlations in $x$-$f$ space and complementary domains.

\begin{table*}[!t]
  \centering
  \caption{Comparison results of different methods on dynamic cardiac cine MRI with high undersampling rate 9 and 12. Best results are indicated in bold.}
  \label{results_2}
  \scalebox{0.95}{
  \setlength{\tabcolsep}{4.5pt}
  \begin{tabular}{ccccccc}  
    \toprule
  \multicolumn{2}{c}{Method} & {$k$-$t$ FOCUSS} & {CRNN-MRI}   & DS+3DCNN  & DS+CRNN &  $k$-$t$ NEXT     \\
  \midrule
  \multicolumn{2}{c}{capacity} & - & 260,866&352,770 & 265,474&374,020\\
  \midrule
\multirow{3}{0.5cm}{$9 \times$} & PSNR & 29.52 (1.58) &32.45 (1.33) &33.47 (1.41) & 33.24 (1.38) & \textbf{34.23} (1.44) \\
& SSIM & 0.951 (0.013) & 0.969 (0.008)& 0.975 (0.006) & 0.975 (0.006) & \textbf{0.979} (0.005)\\
& HFEN & 0.340 (0.033)&0.249 (0.032) &0.214 (0.026) &0.215 (0.027) & \textbf{0.196} (0.030)\\
  \midrule
\multirow{3}{0.5cm}{$12 \times$} & PSNR & 28.14 (1.56) &31.30 (1.32) &32.46 (1.36) & 32.34 (1.35) & \textbf{33.18} (1.40) \\
& SSIM & 0.937 (0.016) & 0.962 (0.009)& 0.969 (0.007) &0.970 (0.007) & \textbf{0.975} (0.005)\\
& HFEN & 0.382 (0.035) & 0.282 (0.034) &0.242 (0.027) &0.239 (0.029) & \textbf{0.225} (0.031)\\
    \bottomrule
  \end{tabular}}
\end{table*}

\begin{figure*}[!t]
\centering
\includegraphics[width=\linewidth]{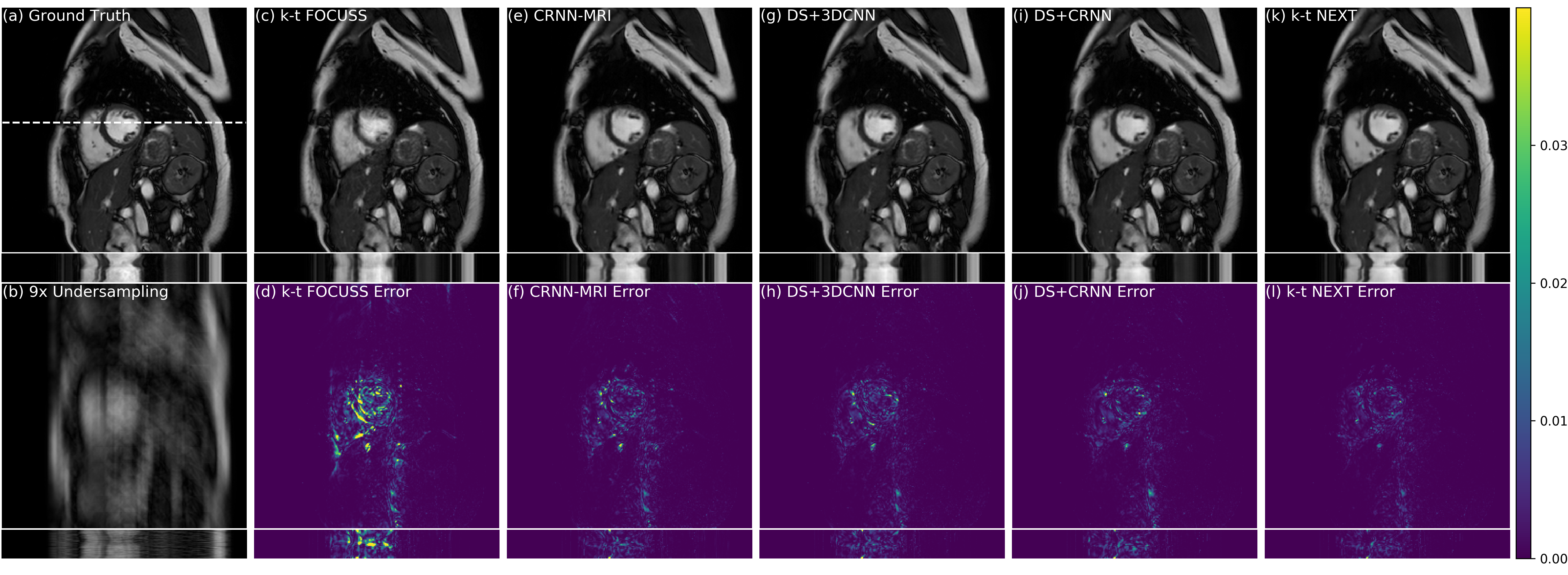}
\caption{Comparison results on spatial and temporal dimensions with their error maps. A dynamic video is shown in supplementary materials for better visualisation.}
\label{fig:xy}
\end{figure*}

\begin{figure*}[!t]
\centering
\includegraphics[width=\linewidth]{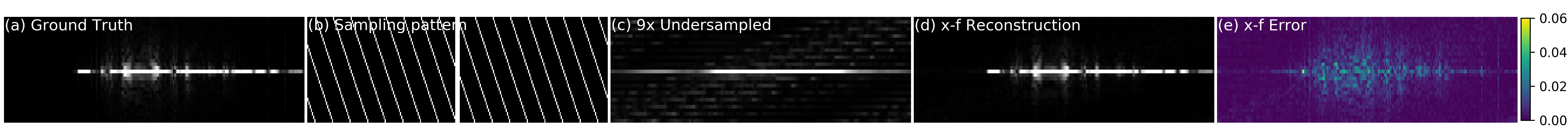}
\caption{Visualisation in $x$-$f$ domain. (a) Ground Truth (b) $k$-$t$ sampling pattern  (c) $9 \times$ undersampled data (d) Reconstructed $x$-$f$ image (e) Error between (c) and (d).}
\label{fig:xf}
\end{figure*}

Additionally, we compared the qualitative results on $9 \times$ undersampled data in Fig. \ref{fig:xy}, where it shows the reconstructed images along both spatial and temporal dimensions, as well as their corresponding error maps. It can be observed that our proposed model can faithfully recover the images with smaller errors especially around dynamic regions compared with other baseline methods. {{In particular, $k$-$t$ NEXT produced visually sharper images than DS methods. This is reflected by the fact that, in contrast to DS approaches which fill in $k$-space data from neighboring frames and therefore could possibly generate averaged and smooth images, $k$-$t$ NEXT directly estimates the missing data in $x$-$f$ space.}} A visualisation of $x$-$f$ reconstruction is also presented in Fig. \ref{fig:xf}, where it displays the reconstructed $x$-$f$ image and its error map in comparison to the input aliased data. It can be observed that the aliasing artefacts were largely removed and the undersampled data were recovered to approximate the ground truth signals.

\section{Conclusion}
In this paper, we have presented a novel deep learning based method, $k$-$t$ NEXT ($k$-$t$ NEtwork with X-$f$ Transform), for highly undersampled dynamic MR image reconstruction. $xf$-CNN is proposed to exploit correlations in $k$-$t$ space via reconstructing the true signals from aliased signals in $x$-$f$ domain. Based on that, $k$-$t$ NEXT is then proposed to learn to iteratively recover the images by alternating between the complementary $x$-$f$ and image domains, where networks from both domains were trained jointly. Experimental results have shown that the proposed $k$-$t$ NEXT outperforms state-of-the-art dynamic MR reconstruction methods in terms of both quantitative and qualitative performance. For the future work, we will extend the method for dynamic 3D applications.
%
%
\section*{Acknowledgements}
This work was supported by EPSRC programme grant SmartHeart (EP/P001009/1).

\bibliographystyle{splncs04}
\bibliography{ref}

\end{document}